\documentclass[preprint,preprintnumbers,amsmath,amssymb,superscriptaddress,floatfix]{revtex4}

\usepackage[dvipdfmx]{graphicx,color}

\usepackage{dcolumn}
\usepackage{bm}
\usepackage{color}
\usepackage{hyperref}
\usepackage{array}

\begin{document}

\title{Fast and selective inter-band transfer of ultracold atoms in bichromatic lattices 
permitting Dirac points}
\noindent
\author{Tomotake Yamakoshi}
\affiliation{Institute for Laser Science, University of Electro-Communications, 1-5-1 Chofugaoka, Chofu-shi, Tokyo 182-8585, Japan}
\author{Shinichi Watanabe}
\affiliation{Department of Engineering Science, University of Electro-Communications, 1-5-1 Chofugaoka, Chofu-shi, Tokyo 182-8585, Japan}

\date{\today}

\begin{abstract}
An experimental group at Beijing~[Yueyang Zhai, {\it et.~al.}, Phys.~Rev.~A {\bf 87}, 063638 (2013)] introduced the method of standing-wave pulse sequence for
efficiently preparing ultracold bosonic atoms into a specific excited band in a 1-dimensional optical lattice. 
Here, we report a theoretical extension of their work to the problem of 1-dimensional bichromatic superlattice.
We find that varying the lattice parameters leads to the so-called Dirac
 point where a pair of excited bands crosses. 
 This paper thus discusses {\it simultaneously} the efficient excitation of the wave packet to the proximity of the Dirac point and its subsequent dynamics in the force field of a parabolic trap.
 With the aid of a toy model, 
 we theoretically unravel the mechanism of the efficient preparation, 
and then numerically explore optimal pulse-sequence parameters for a realistic situation.
 We find an optimized sequence of a bichromatic optical lattice that excites more than 99\% of the atoms 
  to the 1st and 2nd excited bands within 100 $\mu$s without the harmonic trap.
 Our main finding is that the system permitting the Dirac point possesses a region of parameters where
 the excited energy bands become nearly parabolic, conducive to robust coherence and isochronicity.
 We also provide an appropriate data set for future experimentation, including effects of the atom-atom interaction by way of the mean-field nonlinear term.
\end{abstract}

\maketitle

\section{introduction}
\label{sect:Introduction}

Ultracold atoms and molecules in optical lattices have been eagerly investigated over the last 20 years\cite{intro}.
These quantum systems have attracted much attention for their high controllability and accessibility as well as their fascinating quantum effects as exemplified by the artificial gauge fields\cite{gauge}. 
The present paper concerns, in this context,
ultracold atoms in a one-dimensional and highly tunable {\it bichromatic} lattice, which system 
enables us to investigate low-dimensional quantum properties subject to a designed band structure.

Once the bichromatic lattices were experimentally realized~\cite{exp-so} in the beginning of the year 2000, 
there ensued the examination of such phenomena as the Landau-Zener tunneling~\cite{LZ}, 
Bloch oscillations~\cite{BO} and St\"uckelberg interferometry
of ultracold matter waves~\cite{SINT}, {\it etc}.
As for the energy bands, D.~Witthaut {\it et.~al.}~\cite{Dirac-t} theoretically suggested that the 1st and 2nd bands would cross if experimental parameters were properly set so that the dynamics near the crossing could be mapped onto the Dirac equation, hence the coinage of the ``Dirac point''.
This theoretical proposal was experimentally examined by T.~Salger {\it et.~al.}~\cite{Dirac-e}; they demonstrated the Landau-Zener transition at the Dirac point, also known as the Klein tunneling,  subject to the optical dipole trapping and the gravitational potential.
To this day, many groups have studied the bichromatic lattice system in terms of 
quasi-relativistic properties~\cite{Relative}, topological properties~\cite{Topological},  
and also in association with the time-wise lattice system~\cite{timewise} {\it etc}.
Recently, B. Reid {\it et.~al.} reported a theoretical study on manipulation of ultracold atoms in the bichromatic lattice  using the Landau-Zener transition caused by a linear external potential~\cite{Queens}.
Unfortunately, the theory falls short of achieving maximal coherence for shaping the wave packet, 
being based on the Bloch oscillation.

Coherent population transfer onto a specific band is a prerequisite to achieve coherent quantum control over a wide range of Hilbert space.
The Aarhus group\cite{Aarhus} and the Hamburg group\cite{Hamburg} succeeded in coherent manipulation
 by using an amplitude modulation of the optical lattice.
This technique holds the conservation of energy and quasimomentum during an inter-band transition, thus suitable for the coherent wave packet shaping\cite{my-1,my-2,my-3} and the band spectroscopy\cite{spectro}
even though the transfer rate is way below unity.
On the other hand, an experimental group in Beijing demonstrated in 2013\cite{Beijing-1} that
a similar technique called the ``standing-wave pulse sequence'' was extremely efficient.
This technique is very straightforward. It repeatedly turns the optical lattice on and off as pulses 
to a confined Bose-Einstein condensation~(BEC) with appropriate time intervals.
In their first experimental paper, they demonstrated transfer from the ground state 
to the 2nd excited band~(D-band), achieving a superposition of the ground band~(S-band) and 
2nd excited band while vacating 99\% of the ground state.
After the demonstration, this technique was applied to promote the wave packet into the 4th excited band (G-band) to study the dynamics of ultracold atoms in the combination 
of an optical lattice and a harmonic trap~\cite{Beijing-2}.
Very recently, they demonstrated that the pulse sequence is also valid in 2D and 3D systems\cite{Beijing-5}.

In this paper, we study the coherent population transfer by the ``standing-wave pulse sequence'' method in the bichromatic optical lattice. (``Optical lattice '' is abbreviated as OL hereafter.)
Because the Bloch states have an unconventional structure due to the Dirac points,
 the transition selection rule indeed becomes modified by the OL of the second harmonic.
First, we theoretically study the Bloch bands and the transfer selection rule purely in the presence of 
the  bichromatic OL or the monochromatic OL.
We numerically simulate the preparation process 
in a condition realistic enough for experimentation~\cite{Beijing-1} afterward.
We shall show that with the given condition, 
population transfer to the 1st and 2nd bands is attainable up to 90\% within 100${\rm \mu}$s.
The preparation time being too short for environmental noises to cause dephasing, 
the ``standing-wave pulse sequence'' method may as well be considered more reliable for 
setting up the desired wave packet.

We think it appropriate to present specific observables, keeping future experiments in mind. 
To this end, we show the momentum distributions that could be observed
by band mapping~\cite{Beijing-4} after following the post-excitation dynamics for a short while.
We show that in the presence of the Dirac point, the energy dispersion curves of the 1st and 2nd excited bands become nearly parabolic for realistic experimental parameters, thus the wave packet excited
to the neighborhood of the Dirac point proves surprisingly robust and nearly isochronic.
We analyse the wave packet dynamics by mapping it onto a semi-classical Hamiltonian~\cite{my-1,my-2,my-3}.

Generally, the atom-atom interactions via the s-wave scattering are non-negligible,
causing the dephasing of the wave packet.
In the treatment of ultracold atomic systems, the interactions are often represented 
by a non-linear term in the frame work of the mean-field approximation\cite{NLGPE}.
In addition to the dephasing, the strong non-linear term modulates the band structure and Bloch waves\cite{NLBloch}, thus alters  the wave packet dynamics in the OL.
In order to examine this point in the context of this paper, 
we solve the time dependent Gross-Pitaevskii equation~\cite{NLGPE} numerically 
with the nonlinear term inclusive for realistic experimental parameters.

The paper is organized as follows.
Sect.~\ref{sect:system} outlines the theoretical model-system. 
Sect.~\ref{sect:numerics} analyzes numerical results of the excitation process and the subsequent dynamics 
caused by the external harmonic confinement.
The effects of the non-linear term will be also discussed.
Sect.~\ref{sect:conclusions} concludes the paper.

\section{Mathematical Definition and basic features of the System}
\label{sect:system}
According to the experimental papers\cite{Beijing-1,Beijing-2,Beijing-3,Beijing-4}, ultracold atoms are initially loaded onto a 3D harmonic trap.
Then, the 1D OL is turned on to shape the relative amplitudes of the Bloch states (on-duty cycle), and then it is turned off to induce relative phase shifts between bands (off-duty cycle). The net effect
is the desired inter-band transition.
These steps are repeated until the wave function $\psi(t)$ reaches the target state $\psi_{\rm target}$.

Here, we consider dynamics of interacting bosonic atoms in the bichromatic OL by solving the Gross-Pitaevskii equation~\cite{NLGPE}.
Some notations and techniques used in this paper are available in our numerical studies presented
in Ref.'s~\cite{my-1,my-2}. 
The 1D version of the system is described by the time-dependent Hamiltonian
$$H=-\frac{\hbar^2}{2m_{a}}\frac{\partial^2}{\partial x'^2}+ \alpha(t) 
\{ V_1 \sin^2 (k_{r}x') +  V_2 \sin^2 (2 k_{r} x') \} + \frac{1}{2}m_a\omega_0^2 x'^2  
+ g_{1D}N |\psi(x')|^2 $$
where $V_1$ is the height of the optical lattice with the period of $\lambda/2$, $V_2$ is 
with the period of $\lambda/4$, $\alpha(t)$ 
equals 1 during an on-duty cycle, otherwise it is 0, $\omega_0$ is 
the frequency of the harmonic trap, $N$ is the number of total atoms and $g_{1D}$ parametrizes 
the effective atom-atom interaction contracted to one degree of freedom. 
We use recoil energy $E_r=\hbar^2 k_{r}^2/2 m_{a}$ as the unit of energy, recoil momentum $k_{r}=2\pi/\lambda$ as the unit of (quasi-)momentum, lattice constant $\lambda/2$ as the unit of length and rescaled time $t=E_r t'/\hbar$ as the unit of time. Here $\hbar$, $\lambda$ and $m_{a}$ correspond to the Planck constant, laser wave length of the optical lattice, and atomic mass, respectively. 
Rescaling the Hamiltonian, we get
\begin{eqnarray}
H = -\frac{\partial^2}{\partial x^2}+ \alpha(t) \{ s_1 \sin^2 (x) +s_2 \sin^2 (2x) \} + \nu x^2 + g |\psi(x)|^2
\label{eq:re-ham}
\end{eqnarray}
where $x$, $s_1$, $s_2$ and $g$ denote $x=k_{r}x'$, $s_1=V_1/E_r$, $s_2=V_2/E_r$ and $g=g_{1D}N/E_r$ respectively.
According to Ref.~\cite{G1D}, 
$\displaystyle{g_{1D}=\frac{4 \hbar^2 a}{m_{a} a_{\perp}^2}\left( 1- 1.4603\frac{a}{a_{\perp}} \right)^{-1}}$
where $a$, $a_{\perp}=\sqrt{ \frac{2\hbar}{m_{a} \omega_{\perp}} }$ and $\omega_{\perp}$ are 
the s-wave scattering length, effective transverse scattering length, and effective transverse trapping frequency, respectively.
The atom treated here is $^{87}Rb$ whose s-wave scattering length is $a=4.6\times 10^{-9}$m. 
The effective transverse frequency is $\omega_{\perp}\sim 2\pi \times 200 Hz$, 
and $g_{1D}$ ranges from $10^{-5}$ to $1$. 
The other parameters are the same as in Ref.~\cite{Beijing-1}.

So the bichromatic OL part of $H$, namely 
$ H_B = -\frac{\partial^2}{\partial x^2} + \{ s_1 \sin^2 (x) +s_2 \sin^2 (2x) \}$ gives
the unperturbed Bloch states $\{\phi^n_q(x)\}$, namely 
\begin{equation}
 \phi^n_q(x) = e^{iqx} \sum_{K} C_B(n,q,K) e^{2iKx}
\end{equation}
where $q$ is the quasimomentum, $n \in \mathbb {N}$ is the band index, 
$K \in \mathbb {Z}$ is the reciprocal vector index and the coefficient $C_B(n,q,K)$ obtains
by solving the recurrent formula
\begin{eqnarray}
(q+2K)^2 C_B (n,q,K) -s_2 C_B (n,q,K-2)/4 -s_1 C_B (n,q,K-1)/4 &&  \nonumber \\
 \nonumber
-s_1 C_B (n,q,K+1)/4 -s_2 C_B (n,q,K+2)/4 && \\
= (E_q^n-s_1/2-s_2/2) C_B (n,q,K), &&
\label{eq:rec}
\end{eqnarray}
where $E_q^n$ represents the eigenenergy of the Bloch state.

\begin{figure}[htbp]
 \begin{center}
 \includegraphics[width=8cm]{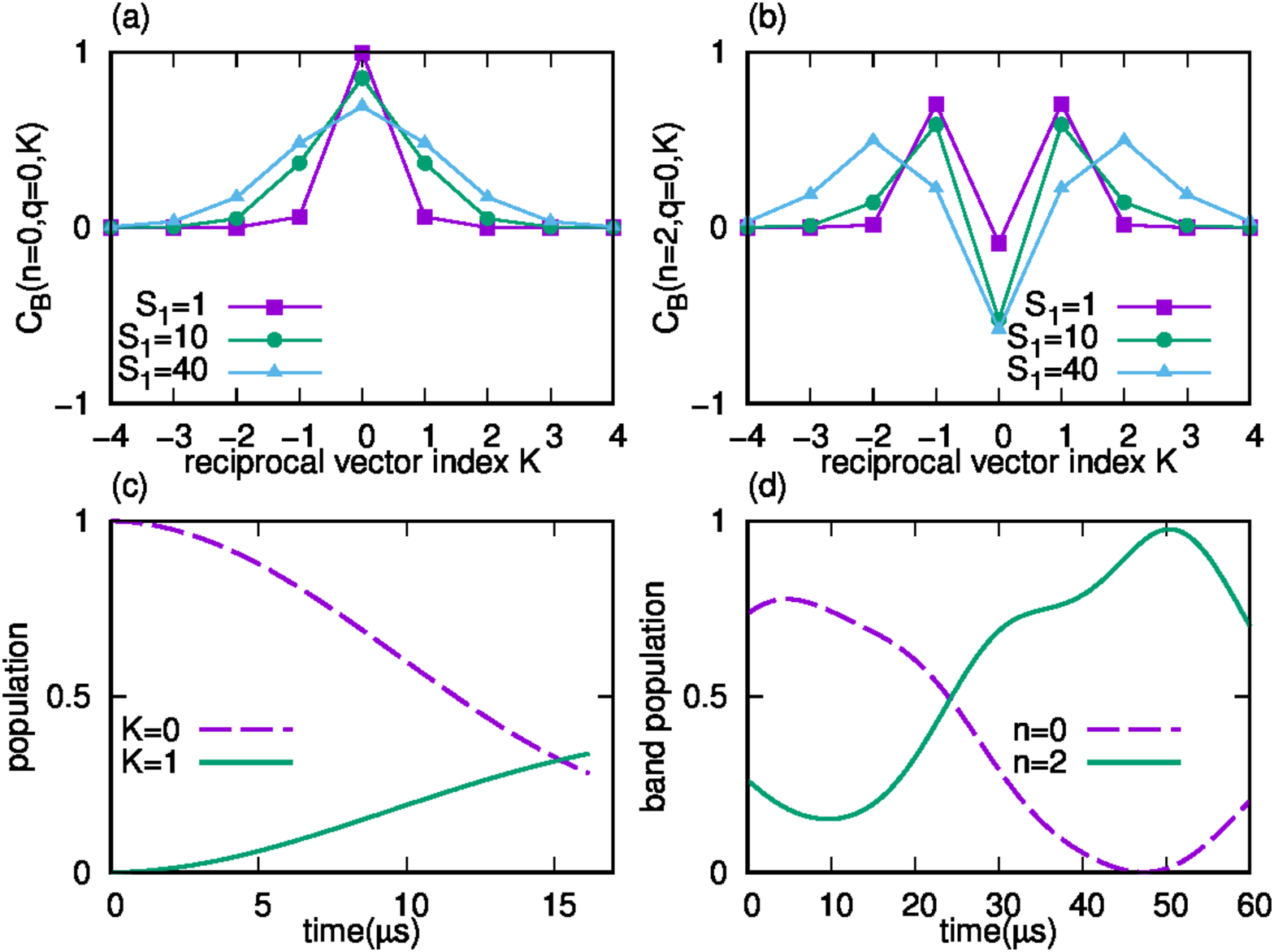}
 \end{center}
 \caption{(Color online) The coefficients $C_B(n,q,K)$ for Bloch states at $q=0$ (a) for the ground band and
  (b) for the 2nd excited band.
 The horizontal axis represents reciprocal vector index $K$.
 Purple squares, green circles and light-blue triangles show the results of 
 $s_1=1$, 10 and 40, respectively.
 Especially, the toy model appropriate to $s_1=10$.
 (c) shows populations of $K=0$ (purple dashed) and $K=\pm1$ (green solid) components 
 during on-duty cycle with $s_1=10$.
 The on-duty process is halted at $t=\tau_1=16.2 \mu s$.
 (d) shows band populations of $n=0$ (purple dashed) and $n=2$ (green solid) components 
 during off-duty cycle with $s_1=10$ after the first on-duty cycle.
 The 2nd band population reaches its maximum around $t=\tau_1^\prime=50.4 \mu s$.
 }
 \label{fig:single-theory}
\end{figure}

In order to grasp the mechanism of inter-band transition in this manipulation, 
we first account for the population transfer to the 2nd band
subject to a single standing-wave pulse sequence.
Consider the following toy-model for a monochromatic OL, namely $s_2=0$ in Eq.~(\ref{eq:rec}),
assuming the initial
wave packet is in the ground state of a very loose parabolic trap
 $\nu\simeq0$. The system is so spread out spatially that the uncertainty principle demands $p_x=0$,
that is $K=0$ with $q=0$ \cite{Beijing-1}.
The model consists of three states, $K=0,\pm1$ with $q=0$ so that
a state vector is represented by a triplet of numbers $(a,b,c)$ such as
$$
(a,b,c)\rightarrow a e^{-2ix} + b + c  e^{2ix}.
$$
The initial state is then
$$
\Psi(t=0)=(0,1,0)
$$
while the normalized eigenvectors $v_n$ $(n=0,1,2)$ are
$$
v_n =(C_B(n,0,-1),C_B(n,0,0),C_B(n,0,+1))
$$
where $n$ is the band index(See Fig.\ref{fig:single-theory} (a) and (b).).
Turning the OL on suddenly is equivalent to projecting 
onto the OL eigen vectors, thus
$$
\Psi(t=0)=\langle v_0|(0,1,0)\rangle v_0 + \langle v_2|(0,1,0)\rangle v_2
=C_B(0,0,0)v_0 +C_B(2,0,0)v_2
$$
Propagating $\Psi$ over the on-duty period $\tau_1$, and then propagating over the off-duty
period $\tau_1^\prime$, we get
$$
  \Psi(t=\tau_1+\tau_1^\prime) =\left(
    \begin{array}{c}
    \{ C_B(0,0,0)C_B(0,0,-1)e^{-iE_0^0\tau_1} + C_B(2,0,0)C_B(2,0,-1)e^{-iE_0^2\tau_1} \} e^{-i4\tau_1^\prime} \\
    C_B(0,0,0)C_B(0,0,0)e^{-iE_0^0\tau_1} + C_B(2,0,0)C_B(2,0,0)e^{-iE_0^2\tau_1} \\
    \{ C_B(0,0,0)C_B(0,0,1)e^{-iE_0^0\tau_1} + C_B(2,0,0)C_B(2,0,1)e^{-iE_0^2\tau_1} \} e^{-i4\tau_1^\prime}
    \end{array}
  \right)
$$
This wave packet becomes proportional to $v_2$ if
\begin{eqnarray*}
f(\tau_1,\tau_1')&=&\left[  C_B(0,0,0) C_B(0,0,1)e^{-iE_0^0\tau_1} + C_B(2,0,0) C_B(2,0,1)e^{-iE_0^2\tau_1}  \right]e^{-i4\tau_1'}C_B(2,0,0)\nonumber \\
                &-& \left[  C_B(0,0,0) C_B(0,0,0)e^{-iE_0^0\tau_1} + C_B(2,0,0) C_B(2,0,0)e^{-iE_0^2\tau_1}  \right]C_B(2,0,1)=0
\end{eqnarray*}

\begin{figure}[htbp]
 \begin{center}
 \includegraphics[width=6cm]{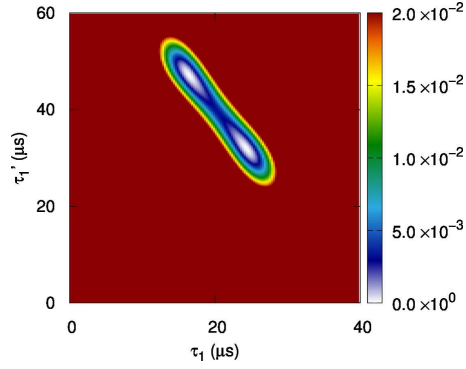}
 \end{center}
 \caption{Doubly-periodic density $|f(\tau_1,\tau_1^\prime)|^2$ 
 over a typical cell with a pair of minima.}
 \label{fig:S10}
\end{figure}
The density $|f(\tau_1,\tau_1^\prime)|^2$ is doubly periodic 
in the present three-state model, thus a typical unit cell appears as in Fig.~\ref{fig:S10}. 
Even under this simple assumption, the 2nd band population is observed to reach 99\% for $s_1=10$ 
as shown in Fig.~\ref{fig:single-theory}(c) and (d).
To summarize, the first on-off duty cycle attains minima of $|f(\tau_1,\tau_1^\prime)|^2$ 
through the phase difference
between the $K=0$ and $K=\pm1$ components, and then in the actual experimental system
the second cycle is applied to make further optimization.

\begin{figure}[htbp]
 \begin{center}
 \includegraphics[width=9cm]{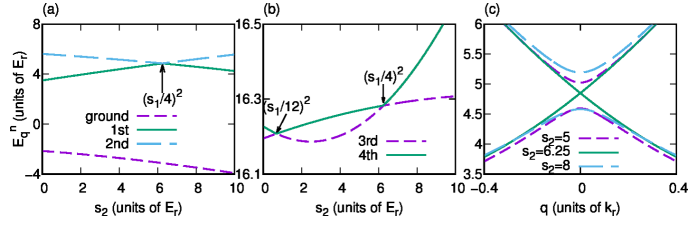}
 \end{center}
 \caption{(Color online) (a) represents ground (purple dashed), 1st (green solid) and 2nd (light-blue long-dashed) band energy as a function of $s_2$ with $q=0$ and $s_1=10$.
1st and 2nd band crosses at $s_2=(s_1/4)^2=6.25$.
(b) represents 3rd (purple dashed) and 4th (green solid) band energy as in (a).
In addition to the crossing at $s_2=(s_1/4)^2$, they cross at $s_2=(s_1/12)^2$. 
(c) represents the lattice height $s_2$ dependence of the band structure around the crossing of 1st and 2nd band.
As shown in (a) the crossing or the Dirac point\cite{Dirac-t,Dirac-e} happens when $s_2=6.25$.
 }
 \label{fig:S10-S2DEP}
\end{figure}

\begin{figure}[htbp]
 \begin{center}
 \includegraphics[width=8cm]{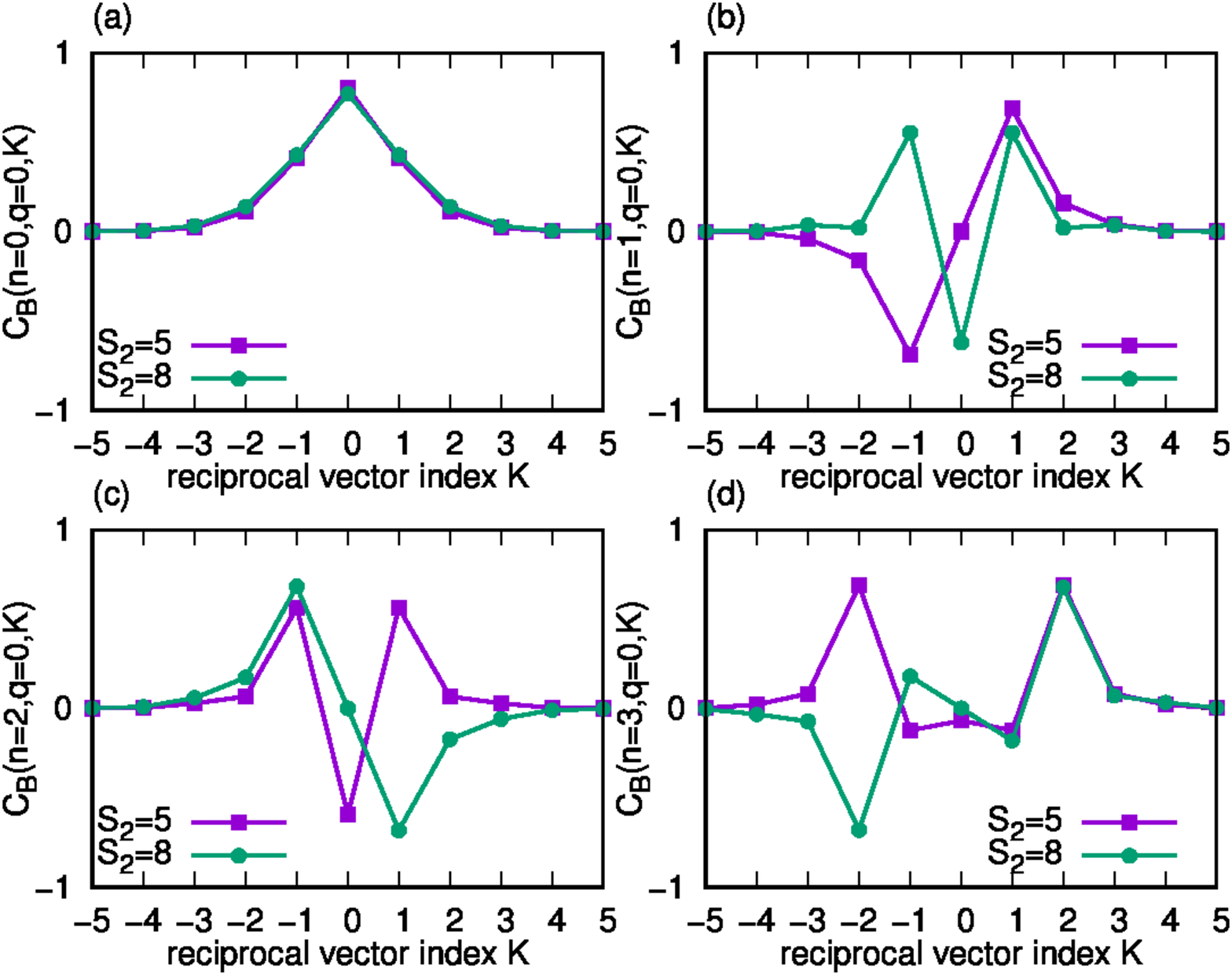}
 \end{center}
 \caption{(Color online) The figures show Bloch coefficients $C_B(n,q,K)$ as a function of $K$ as in
 Fig.~\ref{fig:single-theory} (a) and (b).
 Purple squares show the results of $s_2=5$ and green circles $s_2=8$.
 Panels (a) through (d) correspond to the ground through the 3rd band.
 Note $s_2=5$ of (b) corresponds to $s_2=8$ of (c) up to sign, and likewise for $s_2=8$ of (b)
 and $s_2=5$ of (c).}
 \label{fig:S10-COEFF}
\end{figure}

This simple theory can be certainly applied to the case of the bichromatic OL as well. 
Fig.~\ref{fig:S10-S2DEP} shows the energy structure as a function of $s_2$ for $q=0$ with $s_1$ fixed to 10.
The energy difference between 1st and 2nd bands decreases as $s_2$ is increased until $s_2=(s_1/4)^2$ 
where it vanishes.
We note that the selection rule is modified due to the symmetry of the Bloch states.
The parity of the 1st and 2nd bands changes across this crossing (see Fig.~\ref{fig:S10-COEFF}), 
which prevents the wave packet from reaching the 2nd band from below by this manipulation.
Further details of the energy structure is discussed in Appendix from the view point of Bloch's theorem,
somewhat differently from the previous study~\cite{Dirac-t}.

\section{Discussions on Numerical Results}
\label{sect:numerics}
In the experimental paper of \cite{Beijing-1}, numerically optimized parameters
for a standing-wave pulse sequence were employed.
We extend it to the bichromatic OL system by applying two standing-wave pulses.
Moreover, we take into account effects of the non-linear term incurred by the atom-atom interaction 
and the acceleration due to the harmonic potential.
Our main purpose is three-fold. We study the preparation process systematically, show the effectiveness of the method, and give sets of parameter values that would assist future experimental investigation.
Here, we fix $s_1\equiv10$ and focus on the range $s_2=5-8$ containing the crossing of interest 
 at $s_2=6.25$.
The band population $B_n(t)=\sum_q \left| \langle \phi_q^n  | \psi(t) \rangle \right|^2$ serves as an index of the inter-band transfer.
Below we deal with the initial excitation of the wave packet, its subsequent propagation in a harmonic trap,
and the atom-atom interaction one by one.

\subsection{Excitation by OL pulse sequence}
\label{sect:excitation}

Let us consider excitation of the initial wave packet. 
First, we validate our numerical approach by comparing with the experimental results of 
the monochromatic OL pulse sequence in Ref.~\cite{Beijing-1}. Here, we assume the harmonic trap and
the atom-atom interaction are both absent. Since dephasing becomes pronounced as the elapsed time gets longer,
the proposed excitation method is expected to be more effective if the total elapsed time is shorter.

\begin{table}[h]
\caption{Parameter sets for monochromatic OL listed for $s_1=10$ and 20.
Here $\tau_i$ and $\tau'_i$ represent the on-duty period and off-duty period, respectively, for the
$i$-th cycle $(i=1,2)$, $\tau_{\rm total}=\tau_1+ \tau'_1  + \tau_2  +\tau'_2 $, and $B_2$ is
the 2nd excited band population.
Label {\it a} for this work, and {\it b} for experiment~\cite{Beijing-1}.
Let us note that although the result of 20{\it b} is numerically reproducible, this particular
pulse sequence is not the best choice of $B_2$.}
\begin{tabular}{cc|ccccc|c}
\hline\hline
   $s$ & $\qquad$  &   $\tau_1$   &   $\tau'_1$       &  $\tau_2$    & $\tau'_2$  & $\tau_{\rm total}$ &$B_2(\tau_{\rm total})$  \\
\hline
10 &  {\it a}      &   24.6       &   28.8            &   7.4        &   2.3         & 63.1 & 0.982    \\
 $\qquad$  &  {\it b}      &   24.5       &   28.8            &   8.1        &   2.2         & 63.6 & 0.982    \\
\hline
20 & {\it a}      &   15.0       &   3.3             &   2.0        &   20.7        & 41   & 0.991    \\
   &  {\it b}      &   17.2       &   25              &  12.5        &   1.1         & 55.8 & 0.973    \\
\hline\hline
\end{tabular}
\label{tab:mono}
\end{table}

Even in the case of the bichromatic OL the pulse-sequence excites the atoms primarily 
to the 1st and/or 2nd excited bands because the second optical lattice 
merely interchanges the Bloch coefficients (up to sign) across the Dirac point
 as noted in Fig.~\ref{fig:S10-COEFF}.
In Table~\ref{tab:two-col}, we show
an extension of Table~\ref{tab:mono}
 for the bichromatic OL with $s_1$ fixed to 10.
As discussed in Sect.~\ref{sect:system}, the process cannot excite atoms to the 1st band 
while $s_2 < s_1^2/16$, but tendency changes abruptly across the critical value $s_2=6.25$, that is 
the process fails to excite atoms to the 2nd band once $s_2$ exceeds 6.25.
In all the cases considered, the sequence succeeded in preparing almost 99\% of 
the atoms into the excited bands.

\begin{table}[h]
\caption{Band population for a set of cycle parameters in the bichromatic 
OL with lattice height $s_1$ fixed to 10.}
\begin{tabular}{c|cccc|cc}
\hline\hline
  $s_2$      &   $\tau_1$   &   $\tau'_1$       &  $\tau_2$    & $\tau'_2$  &  $B_1(\tau_{total})$ &  $B_2(\tau_{total})$ \\
\hline
5         &   3.8       &   4.9             &   23.1       &   28.9        &   0.000   &  0.999                                  \\
6         &   21.4       &   30.8             &   8.4       &   9.8        &   0.000   &  0.999                               \\
6.25      &   21.6       &   30.5            &   8.7        &   10.0        &   0.479   & 0.520                                \\
7         &   21.4       &   30.0             &   8.7        &   11.2        &   0.998   & 0.000                                \\
8         &   13.0       &   22.9             &   2.5        &   30.4        &   0.995   &  0.001                               \\
\hline\hline
\end{tabular}
\label{tab:two-col}
\end{table}

\subsection{Effects of the harmonic potential}
\label{sect:harmonic}
In the Beijing experiment, an atomic BEC is set up in a 3D harmonic trap first.
Even in the limit of non-interacting atoms, 
the initial wave packet then has a Gaussian distribution of finite width in momentum space.
Strictly speaking, the harmonic trap accelerates the atoms during the excitation process 
so that it is advisable to turn off the trap during the pulse sequence as in Ref.~\cite{Beijing-1}.
However, in this paper we consider the pulse sequence with the harmonic trap on in order to check its effect 
on the post-excitation dynamics of the wave packet~\cite{Beijing-4}.

\begin{figure}[htbp]
 \begin{center}
 \includegraphics[width=7cm]{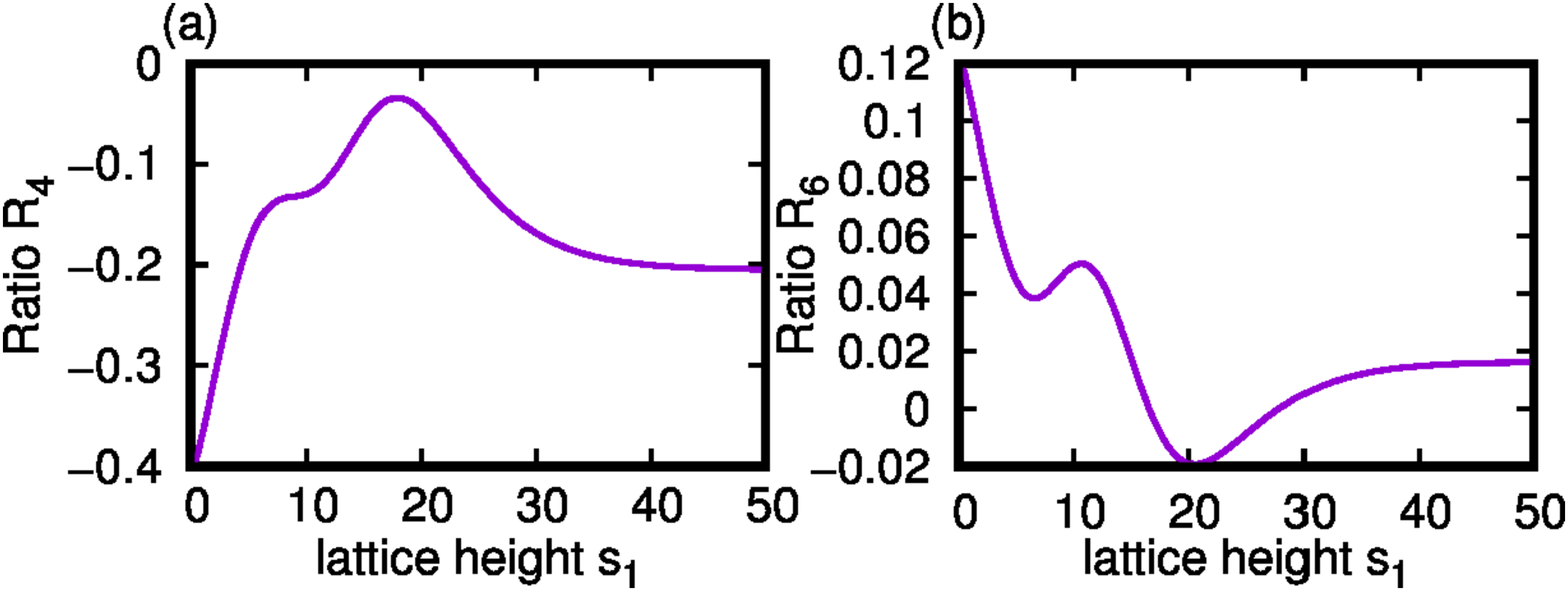}
 \end{center}
 \caption{(Color online) 
 The ratio of (a) quartic term to quadratic 
 $R_4=\frac{\pi^2}{48} \frac{ \sum_{m=1}^\infty m^4 J_m }{\sum_{m=1}^\infty m^2 J_m} $  and (b) sextic to quadratic $R_6= \frac{\pi^4}{5760} \frac{ \sum_{m=1}^\infty m^6 J_m }{\sum_{m=1}^\infty m^2 J_m} $.
 In the limit of $s_1\gg1$, $|R_4|$ goes to 0.206 and $|R_6|$ goes to 0.017, which are rather small.
 }
 \label{fig:ratio}
\end{figure}

In order to study the dynamics in quasimomentum space, experiments apply the band-mapping technique
to the excited components by varying the hold time while keeping the OL and 
harmonic potential both on~\cite{Aarhus,Hamburg,Beijing-4}. Thus we examine the dynamics of the excited wave packet
in quasimomentum space numerically as well as semi-classically 
following the Hamburg group \cite{Hamburg,my-1,my-2}.
According to the semi-classical theory, the wave packet obeys 
\begin{eqnarray}
H_{cl} = E_q^n + \nu x^2,
\label{eq:cl-ham}
\end{eqnarray}
under the single-band approximation.
This semi-classical Hamiltonian allows us to calculate the outermost location of the excited 
wave packet $x_{\rm max}$ and the critical time $\tau_c$ equaling how long it takes
 the wave packet to reach the edge $q_0$ of the excited band where $q_0$ equals 0 
for even-indexed bands and 1 for odd-indexed bands.
Here we assume that the excited wave packet is initially located at $(x_i,q_i)=(0,\nu^{1/4}/\sqrt{2})$
where $q_i$ corresponds to the variance of the ground state wave function in momentum space, 
and that the shape does not change during the pulse sequence.
The semiclassical expressions are
$x_{\rm max}=\pm \sqrt{\frac{E_{q_i}^n - E_{q_0}^n }{\nu}}$ 
and
$\tau_c = \frac{1}{2\sqrt{\nu}}\int_{q_0}^{q_i}\left( \sqrt{{E_{q_i}^n - E_{q}^n }} \right)^{-1} dq$.

This single-band approximation breaks down 
when the Dirac point appears as in Fig.~\ref{fig:S10-S2DEP}(c).
To gain a better understanding, let us think what would happen to 
the lowest band of the OL of period $\lambda/4$ if the OL of period $\lambda/2$ were slowly turned on?
First, it would break into two bands with a gap at the zone boundary of the $\lambda/2$-OL with an
additional band eventually settling down from above as the true lowest band. 
But the two originally disconnected bands become degenerate at the particular coupling strength induced by the presence of the new lowest band, a phenomenon frequently encountered in a three-level system.
The energy dispersion curve appears to restore the feature of the OL of period $\lambda/4$ at the Dirac point, 
but the content of each eigenvector changes across it, reflecting the  period of $\lambda/2$. 
The motion of the wave packet in the region covering the Dirac point is best analyzed with the aid of the extended zone representation suitable for the $\lambda/4$-OL. 
The band dispersion is thus given 
by $E(q_{\rm ext})=A - \sum_{m=1}^\infty 2 J_{m}\cos(m \pi q_{\rm ext}/2)$ 
for the 1st($|q_{\rm ext}| \leq 1 $) and 2nd($1 \leq |q_{\rm ext}| \leq 2$) excited bands where A and $J_{m}$ are 
the energy offset and the $m$-th order hopping constant, respectively.
The energy dispersion could be expanded into Taylor series as 
$E(q_{\rm ext})= A - 2 \sum_{m=1}^\infty J_m + 
\frac{\pi^2 q_{\rm ext}^2}{4} \sum_{m=1}^\infty m^2 J_m - 
\frac{\pi^4 q_{\rm ext}^4}{192} \sum_{m=1}^\infty m^4 J_m + \cdots$.
In the limit of $s_1\gg1$ while holding $s_2=(s_1/4)^2$, the dispersion reduces to a single cosine function. 
In contrast, it goes to a superposition of cosine terms in the limit of $s_1\ll 1$,
as is known by the tight-binding model~\cite{SSP}.
As shown in Fig.~\ref{fig:ratio}, there is a region between these two limits 
where the dispersion approaches a parabolic function due to the destructive interference of cosine terms.
In the case of $s_1=10$, this effect suppresses the higher-order terms 
so that the classical Hamiltonian can be well approximated by 
$H_{cl-ext} = \sum_{m=1}^\infty \pi^2 m^2 J_m  q_{\rm ext}^2/2 + \nu x^2$, 
making the excited wave packet robust. The wave packet thus enjoys isochronicity to
an unexpectedly high degree.
Moreover, the dispersion near the Dirac point  is 
approximately given by $E(q)=A-\pi J_{1} q$ in the limit of $s_1\gg1$ where $q=q_{\rm ext}\pm1\sim0$
is quasimomentum in $E(q_{\rm ext})$ above in the {\it reduced zone} representation. 
This dispersion is thus in line with the explanation by the Dirac equation~\cite{Dirac-t,Dirac-e}.

\begin{figure}[htbp]
 \begin{center}
 \includegraphics[width=7cm]{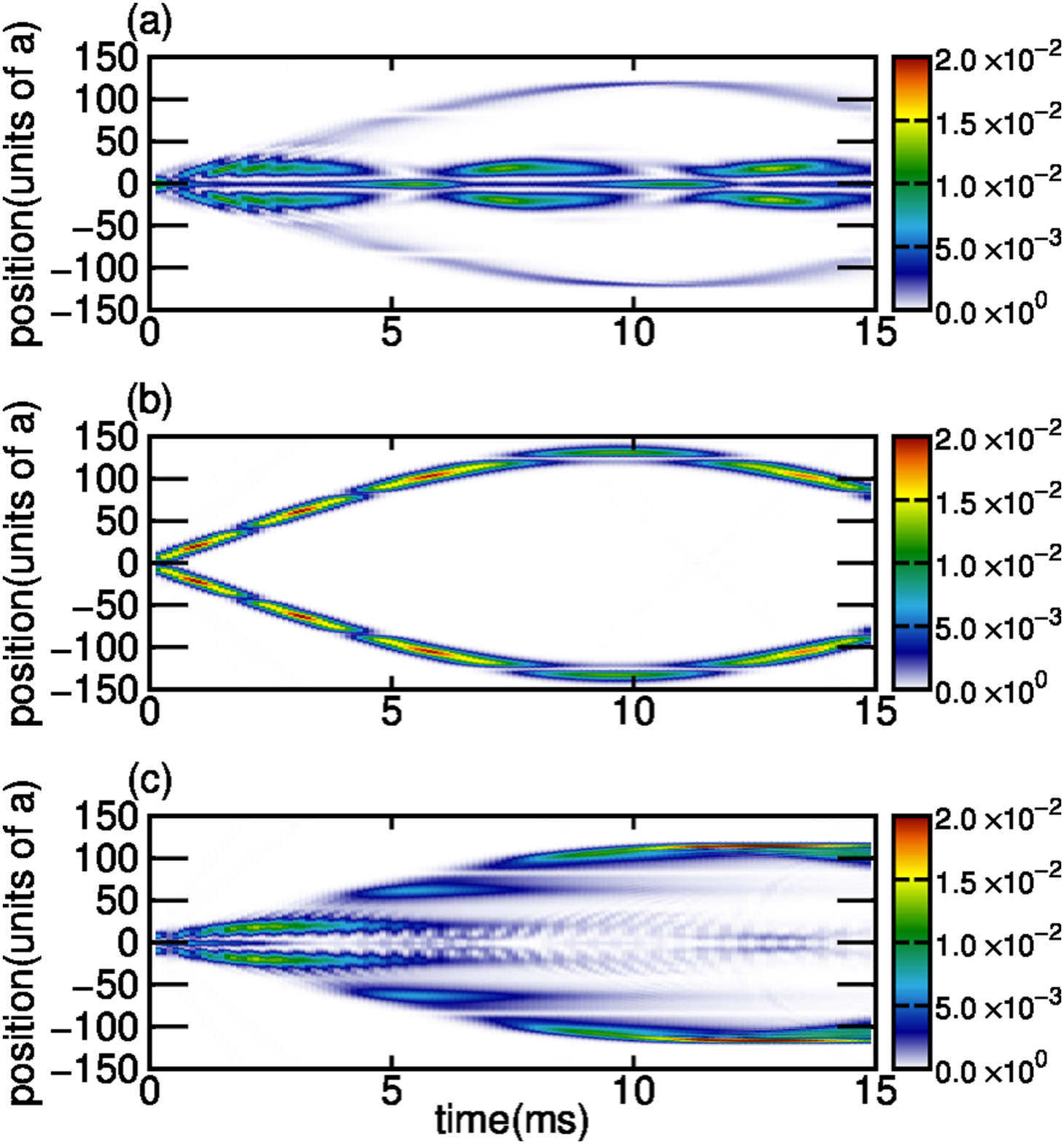}
 \end{center}
 \caption{(Color online) The figures show time-evolution of the wave packet in position space as a function of hold time with $s_1=10$ and $\nu=1.0\times10^{-5}$.
 Time $t=0$ corresponds to immediately after the 2nd off-duty cycle.
 Density becomes denser toward red and lower toward white.
 (a) In the case of $s_2=5$, the excited wave packet is mostly located around the bottom of the 2nd band, thus it is located around the origin of the harmonic potential.
 (b) In the case of $s_2=6.25$, the band gap becomes closed, therefore the excited wave packet smoothly traces the band structure.
 This results in a less dispersive motion of the wave packet in position space.
 (c) In the case of $s_2=8$, the excited wave packet gets located around the top of the 1st band, therefore it traces the 1st energy-band due to the harmonic potential.  The wave packet can thus travel far away from the origin. 
 }
 \label{fig:20HZ-HOLD}
\end{figure}

\begin{figure}[htbp]
 \begin{center}
 \includegraphics[width=9cm]{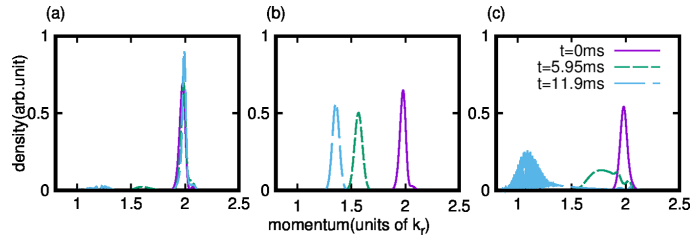}
 \end{center}
 \caption{(Color online) The momentum distributions after the band mapping procedure with hold time 
 $t_{\rm hold}=$ 0 (purple solid), 5.95 (green dashed) and 11.9 (light-blue long-dashed) ms 
 in the case of $\nu=1.0\times10^{-5}$.
 (a) corresponds to $s_2=5$, (b) to $s_2=6.25$ and (c) to $s_2=8$.
 }
 \label{fig:20HZ-MOMENTUM}
\end{figure}

Let us look at the numerical results of the quantum simulation.
Fig.~\ref{fig:20HZ-HOLD} shows the post-excitation evolution of the wave packet 
during the hold time.
Here the parameters are $\nu=1.0\times10^{-5}$ ($2\pi\times 20$ Hz in SI unit), $s_1=10$, and $s_2=5$ for (a), $s_2=6.25$ for (b), and $s_2=8$ for (c), respectively.
In the case of (a), most of the atoms are transferred to the bottom of the 2nd band, and 
only a small fraction of the wave packet goes to the top of the 1st band. 
In addition, the Landau-Zener transition between 1st and 2nd bands is negligible\cite{my-1,Aarhus2}. 
The atoms are thus trapped in the 2nd band.
In the case of (b), the gap is closed. 
Therefore, the wave packet traces the band structure without reflection at the edge of the 2nd band.
This results in the well-defined sharp wave packet in position space.
In the case of (c), the wave packet traces the 1st band, and their typical wave packet motion is 
characterized by $x_{\rm max}= 117$ lattice sites and $\tau_c = 11.9$ ms.
We note that all the figures show some beats, that is characteristic interference patterns.
This effect is due to the non-parabolic dispersion of the band structure and 
the broadening of the initial wave packet in momentum space.

\begin{figure}[htbp]
 \begin{center}
 \includegraphics[width=5cm]{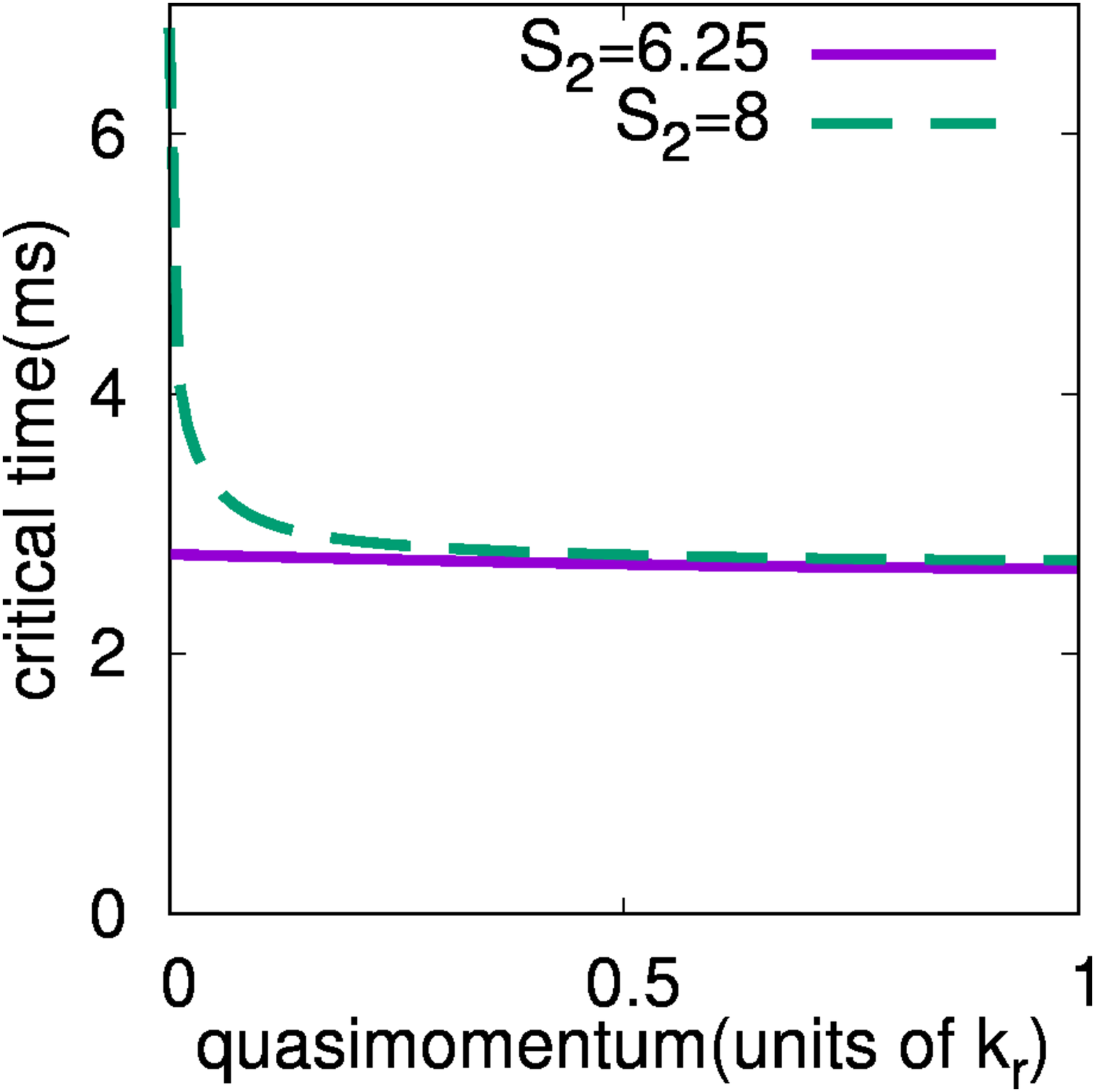}
 \end{center}
 \caption{(Color online) Critical time $\tau_c$ as a function of quasimomentum in the 1st excited band.
 Purple solid and green dashed lines correspond to $s_2=6.25$ and 8.
 In the case of $s_2=6.25$, the critical time is almost independent of quasimomentum. 
 However, in the case of $s_2=8$, the critical time diverges at $q=0$. 
 See text for detailed discussion.}
 \label{fig:critical-time}
\end{figure}

Fig.~\ref{fig:20HZ-MOMENTUM} shows momentum distributions after the band mapping procedure effected
at representative values of hold time. 
(Note we plot only the positive part of the momentum space 
because the density distribution always keeps its symmetric feature with respect to $q=0$.)
We employ the same band mapping procedure as in the experimental paper \cite{Beijing-4}, 
namely, we fix the time duration equal to 1 ms, 
and then let the lattice height decay with the decay time constant equal to 100 $\mu$s.
The resulting distribution is what would be observed experimentally.
As we discussed above, the momentum distribution is almost localized around $p=2$ in (a).
However, in (b) and (c) the main part of the momentum distribution travels in the first band $1<p<2$.
Especially, in the case of (b), the wave packet appears 
less dispersive than the other two cases since the classical Hamiltonian 
is well approximated by that of the 1-dimensional harmonic trap.
In order to confirm this feature, we plot the critical time $\tau_c$ 
as a function of the initial quasimomentum in Fig.~\ref{fig:critical-time}.
In the case of $s_2=6.25$, the distribution is almost flat 
just like the harmonically trapped system in the absence of the OL.

Incidentally, let us allude to the relationship between the present problem and that of the classical non-linear pendulum.
By using the band dispersion, {\it i.~e.}~$E(q_{\rm ext})=A - J_{1}\cos(\pi q_{\rm ext}/2)$ valid 
in the limit of high lattice height, 
the critical time is given by the complete elliptic integral of the first kind~\cite{math},
 {\it i.~e.}~$\tau_c(q_i)=\frac{1}{\pi \sqrt{J_1\nu}}K[\sin^2 \left\{ \pi(1-q_i)/4 \right\}]$.
The difference between $q_i=0$ and $1$ is about  18\% according to this equation. 
However, a fully numerical simulation with $s_1=10$ yields an even smaller difference of 4\%.
On the other hand, in the case of $s_2=8$, the critical time $\tau_c$ 
diverges at $q_i=0$ due to one unstable saddle point in phase space, 
an analogue of the hyperbolic point of the pendulum, 
causing the wave packet to diffuse both in position and momentum space.

\begin{figure}[htbp]
 \begin{center}
 \includegraphics[width=9cm]{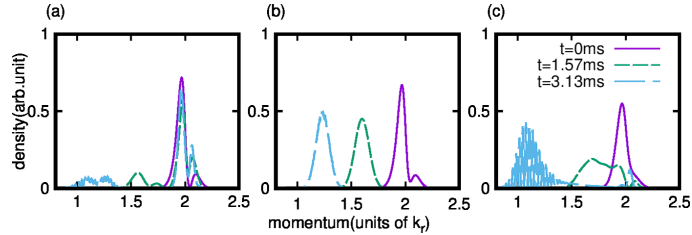}
 \end{center}
 \caption{(Color online) Same as Fig.~\ref{fig:20HZ-MOMENTUM} for $\nu=1.2\times10^{-4}$.
 The hold time $t_{\rm hold}=$ 0, 1.57 and 3.13 ms }
 \label{fig:70HZ-MOMENTUM}
\end{figure}

Let us also examine the dynamics with tighter harmonic trap $\nu=1.2\times10^{-4}$ ($2\pi\times 70$ Hz in SI unit).
In this case, the initial wave function has a broader momentum distribution than in the previous 
looser case.
The features are almost the same as the case of $\nu=1.0\times10^{-5}$, however, the tighter trap reduces 
both the time scale of the periodic motion and the length scale of the position space.
In the case of (c), the characteristic values are $x_{\rm max}= 33$ lattice sites and $\tau_c = 3.13$ ms.
Fig.~\ref{fig:70HZ-MOMENTUM} also shows the momentum distribution after the band mapping.
In comparison to Fig.~\ref{fig:20HZ-MOMENTUM}, each momentum distribution shows a broader shape, 
reflecting the spatial tightness of the initial wave function. 

\begin{figure}[htbp]
 \begin{center}
 \includegraphics[width=9cm]{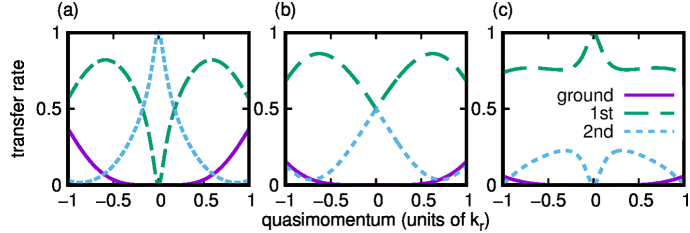}
 \end{center}
 \caption{(Color online) Population transfer rate as a function of quasimomentum 
 with the time sequence shown in Table~\ref{tab:two-col}.
 (a), (b) and (c) correspond to $s_2=5$, 6.25 and 8.
 Purple solid, green long dashed and light-blue short dashed lines show ground, 
 1st and 2nd band, respectively.
 In the case of (a), the distributions has sharp peak (kink) for 2nd (1st).
 As the lattice height $s_2$ increases toward (c),
 the peak and the kink in (a) flip over, making the distribution seemingly less $q$-dependent.
 See text for more details.}
 \label{fig:pops-quasi}
\end{figure}

How does the population transfer rate depend on the trap strength?
To see this, we plot the population transfer rate $R_0(q)$ calculated without a trap 
in Fig.~\ref{fig:pops-quasi} as a function of quasimomentum 
using the time intervals shown in Table~\ref{tab:two-col}.
The shape of the excited wave packet at the pulse sequence's end could be estimated 
by this transfer rate $R_0(q)$  times the initial quasimomentum distribution integrated over $q$.
As shown in Fig.~\ref{fig:pops-quasi}, $R_0(q)$ being close to 1 around $q=0$,
 the initial wave packet localized at $q=0$ would be ideal for selective momentum transfer.
In Table~\ref{tab:band-pop}, we show the 1st and 2nd band population at the pulse end 
with $\nu=1.0\times10^{-5}$ and $\nu=1.2\times10^{-4}$.
These representative data suggest that the harmonic trap is nonnegligile and tends to make 
the excited wave packet more delocalized in momentum space. 
Because the transfer rate $R_0(q)$ has non-uniform distribution, the population transfer rate
gets reduced by the harmonic trap. 
For instance, in Table~\ref{tab:band-pop} as we look at the $n=2$ component, 
it is 91\% at $\nu=1.0\times10^{-5}$ 
whereas it is 81\% at $\nu=1.2\times10^{-4}$ (Fig.~\ref{fig:pops-quasi}(a) and (c)).
Thus, the pulse sequence with a tight harmonic trap may not be advisable for selective momentum transfer.

Another point is that although
numerical optimization with a more appropriate initial condition in a harmonic trap 
would doubtlessly give more efficient parameters, it would lead to dealing with a huge number of
simultaneous equations since evaluation of the population transfer rate is done $q$ by $q$.
The tighter the trap, the more equations needed.
There would be limit to the number of simultaneous equations handleable 
in actual numerical optimization.

Furthermore, Fig.~\ref{fig:pops-quasi} shows that the transfer rate becomes flat as $s_2$ increases.
This result indicates that the excitation process with high $s_2$ makes it 
less $q$-dependent.
Indeed as shown in Table~\ref{tab:band-pop}, the reduction rate is 0.808/0.912=0.886
 for the 2nd band with $s_2=5$, and 0.897/0.953=0.941 for the 1st band with $s_2=8$.
As far as excitation goes, 
the wave packet's acceleration by the harmonic trap is unimportant
since the duration of 100$\mu$s is much shorter than the period of several milliseconds
of motion in the trap.

\begin{figure}[htbp]
 \begin{center}
 \includegraphics[width=7cm]{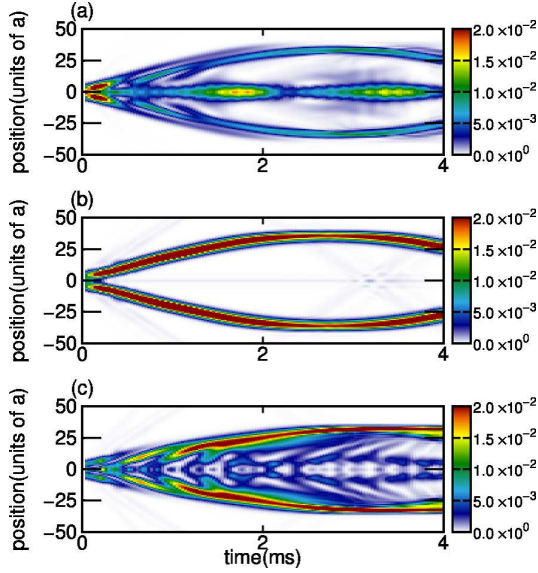}
 \end{center}
 \caption{(Color online) Time-evolution of the wave packet in position space as a function of hold time. 
 Same as Fig.~\ref{fig:20HZ-HOLD} with $\nu=1.2\times10^{-4}$.
 }
 \label{fig:70HZ-HOLD}
\end{figure}

\begin{table}[h]
\caption{1st and 2nd band population at the end of the pulse sequence with $s_1=10$.}
\begin{tabular}{c||c|c||c|c||c|c}
\hline\hline
 & \multicolumn{2}{c||}{$s_2=5$} & \multicolumn{2}{c||}{$s_2=6.25$}  & \multicolumn{2}{c}{$s_2=8$} \\
    \cline{2-7}
$\nu$    &   $n=1$   &   $n=2$       &  $n=1$    & $n=2$  &  $n=1$ &  $n=2$ \\
\hline
 $1.0\times10^{-5}$   &   0.084   &   0.912     &   0.550    &   0.443   &   0.953   &  0.037    \\
 $1.2\times10^{-4}$   &   0.186   &   0.808     &   0.585    &   0.400   &   0.897   &  0.088    \\
\hline\hline
\end{tabular}
\label{tab:band-pop}
\end{table}

\subsection{Effects of the atom-atom interaction}
\label{sect:non-linear}

One of the reasons why ultracold atomic systems are considered to 
offer a fascinating experimental playground is 
that the strength of the non-linear interaction is controllable by the Feshbach resonance~\cite{Feshbach}
so that the strongly interacting regime becomes experimentally accessible with ease.
According to previous studies, the atom-atom interaction alters the conventional band structure,
 causing, for instance, the so-called non-linear Bloch bands
presenting loop-like structures at the band edge~\cite{NLGPE},
 solitary wave packets~\cite{NLBloch} and so forth. 
Such changes show up when the strength of the non-linear term becomes comparable to the lattice height\cite{NLsuper}.
In the experimental system considered, the non-linear term is typically 
quite small like $10^{-5}-10^{-3}$. Nevertheless, in order to make the influence of the
non-linear term clearly visible, we extend its range from 0 up to 1 in this paper.

\begin{figure}[htbp]
 \begin{center}
 \includegraphics[width=7cm]{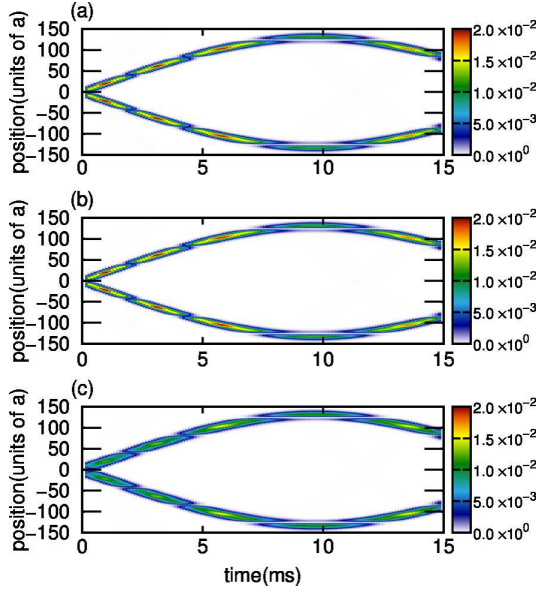}
 \end{center}
 \caption{(Color online) Time-evolution of the isochronic excited wave packet 
 in position space as a function of hold time 
 with (a)$g=5\times10^{-4}$, (b)$1\times10^{-2}$, and (c)1.
 Here $s_1=10$, $s_2=6.25$ and $\nu=1.0\times10^{-5}$.
 All the cases show clear and robust oscillation as in Fig.~\ref{fig:20HZ-HOLD}(b).
 }
 \label{fig:NONL-20HZ-HOLD}
\end{figure}

\begin{figure}[htbp]
 \begin{center}
 \includegraphics[width=9cm]{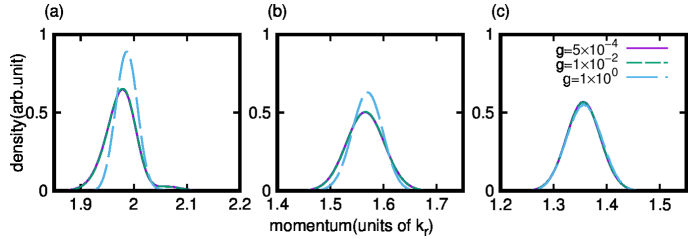}
 \end{center}
 \caption{(Color online) Momentum distributions after the band mapping process with hold time (a)t=0, (b)5.95ms and (c)11.9ms.
 See text for details.
 }
 \label{fig:NONL-20HZ-MOMENTUM}
\end{figure}

Fig.~\ref{fig:NONL-20HZ-HOLD} shows the time-evolution of the {\it isochronic} wave packet 
after the pulse sequence at various values of the effective interaction $g$.
For this reason, we use the same parameter set as for Fig.~\ref{fig:20HZ-HOLD}(b), 
but with (a) $g=5\times10^{-4}$, (b) $1\times10^{-2}$, and (c) 1.
Even in the case of (c) with $g=1$, there is no dramatic change; therefore the system is insensitive to the interaction strength in a practical parameter regime.
Nevertheless, the non-linear term broadens the initial wave packet in space to 
an extent noticeable by scrutiny so that 
the excited wave packet in (c) is also slightly broadened in space.
On the contrary, in momentum space, the non-linear term contracts the initial wave packet, 
therefore the excited wave packet immediately after the pulse sequence shows a narrower distribution 
as shown in Fig.~\ref{fig:NONL-20HZ-MOMENTUM}(a).
However, the wave packet with $g=1$ spreads out gradually in momentum space
due to the interaction which imparts momentum (Fig.~\ref{fig:NONL-20HZ-MOMENTUM}(b)),
thus catching up with the cases with smaller values of $g$, and the distributions thus coincide after 11.9~ms.
The semi-classical treatment then becomes adequate. 
This result indicates that the non-parabolic dispersion dominates 
the early dynamics in the parabolic lattice with $\nu=1.0\times10^{-5}$. 
At any rate, it appears difficult for real experiments to reveal clear indication of the non-linearity.   

\section{Conclusions}
\label{sect:conclusions}

Our simple toy model based on Bloch's theorem added a somewhat more comprehensible picture 
 to the excitation by the standing-wave pulse sequence~\cite{Beijing-1} than previously given.
We have theoretically examined the extension of the method to
 the bichromatic superlattice, and observed an unconventional crossing, namely the Dirac point, 
 between the 1st and 2nd excited bands. 
The excitation selection rule is modified near the Dirac point accordingly, 
depending on the height of the second optical lattice.
Our numerical results show that the standing-wave pulse sequence is valid for the bichromatic optical superlattice.
 In an ideal case, the population transfer from the ground band to the 1st and 2nd bands
  is shown to be attainable with 99\% efficiency and within 100$\mu$s.

In addition, we numerically examined the effects of the harmonic trap and the atom-atom interaction.
We mainly focused on the dynamics of the excited wave packet after the pulse sequence, 
at times including the decay of the applied optical lattice 
to mimic the experimental band mapping technique.
This technique is used for identifying momentum distributions of the excited wave packet.

The classical theory accounts for the bulk of the dynamics in phase space.
We have seen that the band-dispersion dominates the dynamics.
Especially, we found that the band dispersion approaches a parabolic curve 
in the presence of the Dirac point particularly with $s_1=10$.
Consequently, the wave packet preparation in the presence of the Dirac point produces 
amazingly robust wave packets as shown in Fig.~\ref{fig:20HZ-HOLD}(b) and Fig.~\ref{fig:70HZ-HOLD}(b).
On account of this inspiring feature, we produced the momentum distribution 
of the isochronic wave packet as a reference for future experiments.

The harmonic potential affects the width of the initial distribution in momentum space, which in turn
 causes a reduction of the transfer rate.
This fact suggests that preparing the initial BEC with a low-frequency harmonic trap would be preferable to
with a high-frequency one.
We also examined the effect of the atom-atom interaction in the framework of the mean-field theory 
which is suitable for the setup of practical experiments. 
We found that the interaction does not substantially alter the dynamics. 
Since we merely consider the 1-dimensional system in this paper, 
it remains to explore how the dimensionality of the system affects the excitation process.
However, its effect would be negligible in practical situations according to our previous study~\cite{my-2} as long as the non-linear term is relatively small in comparison to the lattice height.

In comparison to the excitation process with the amplitude modulation~\cite{Aarhus,Hamburg}, 
the pulse-sequence excites atoms into higher bands in a shorter period of time and with higher efficiency
than other methods.
Consequently the combination of the pulse-sequence and the acceleration by an external potential
 would be useful for population transfer 
at a particular instant, starting with the wave packet initially localized in momentum space.
Especially, creating solitary wave packets in the presence of the Dirac point 
would be a fascinating application. Such an attempt is akin 
to optical soliton generations in the field of quantum optics.
Another fascinating application is an investigation of the topological dynamics 
in higher bands such as the topological pumping~\cite{pump}.
However, the pulse-sequence may not be suitable for the selective momentum transfer 
such as the hole creation in Fermionic quantum 
degenerate gases~\cite{Hamburg,spectro} since it violates 
the energy conservation law.
This fact suggests that a suitable combination of the pulse-sequence and 
the amplitude modulation would lead to a powerful strategy for precise and 
coherent manipulation of the atomic wave packet subject to given energy band structures.

\section*{Acknowledgments}
This work was supported by Research and Educational Consortium for Innovation of Advanced Integrated Science(CIAiS) and JSPS KAKENHI Grant Number 17K05596.
We thank Mr.~T.~Hosaka for his help with generating some preliminary numerical data set.

\appendix
\section{Energy bands of an atom in the bichromatic lattice}
\label{sect:band-structure}
In this section, we present a simple treatment of the band structure of the 1-dimensional bichromatic lattice.
Here we apply the second-order perturbation theory, regarding the second harmonic as
the perturber to the monochromatic OL, 
and explain the crossing features of the bichromatic band structure discussed in Sec.~\ref{sect:system}.
The Hamiltonian 
$H_{B1} = -\frac{\partial^2}{\partial x^2} + s_1 \sin^2 (x)$ gives the Bloch states of 
the non-interacting bosonic atoms.
Each Bloch state is represented as 
\begin{eqnarray}
\chi^n_q(x) = e^{iqx} \sum_{K} C_{B1}(n,q,K) e^{2iKx}
\label{eq:bloch-single}
\end{eqnarray}
where the coefficient $C_{B1}(n,q,K)$ derives from 
the recurrent formula,
\begin{eqnarray}
(q+2K)^2 C_{B1} (n,q,K) -s_1 C_{B1} (n,q,K-1)/4&& \nonumber \\ 
 -s_1 C_{B1} (n,q,K+1)/4 &=& (e_q^n -s_1/2)C_{B1} (n,q,K).
\label{eq:rec-single}
\end{eqnarray}

\begin{figure}[htbp]
 \begin{center}
 \includegraphics[width=5cm]{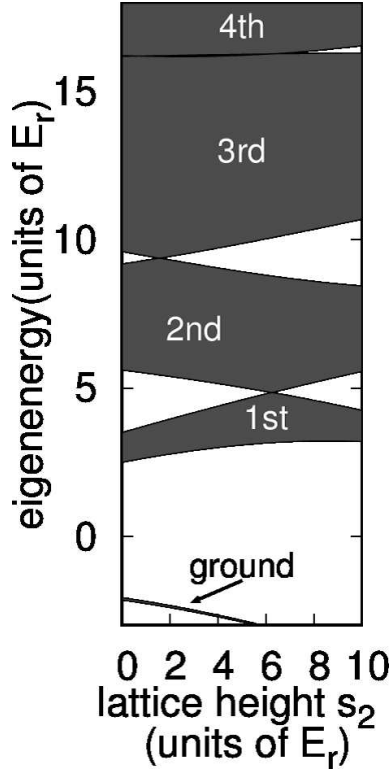}
 \end{center}
 \caption{(Color online)The stability diagram of the bichromatic lattice with $s_1=10$.
 See text for details.
 }
 \label{fig:strutt}
\end{figure}

Fig.~\ref{fig:strutt} shows the stability diagram known as the Strutt diagram for the bichromatic OL system in the plane of $E_q^n$ of Eq.~(\ref{eq:rec}) versus $s_2$ with $s_1$ fixed to 10. 
A similar stability diagram is familiar
in the problem of the Mathieu equation~\cite{math}.
Shaded regions pertain to the stable solutions, and the counterparts to the unstable ones.
As mentioned in Sec.~\ref{sect:system}, the Dirac points appear at the intersections of the
shaded regions, for instance at $s_2=(s_1/4)^2=6.25$ for the bands labeled 1st and 2nd, and 
at $(s_1/12)^2\simeq0.694$ for 2nd and 3rd.
An additional Dirac point appears at $s_2=(s_1/8)^2\simeq1.56$ corresponding to $q=\pm1$.
The pulse sequence transfers only those atoms near $q=0$, therefore we focus on $q=0$ hereafter.

The second-order perturbation theory yields the eigenenergies of the bichromatic OL system,
\begin{eqnarray}
\tilde{E}_q^n  \simeq e_q^n +\frac{s_2}{2} - \frac{s_2}{4} d_n + \frac{s_2^2}{16} f_n,
\label{eq:pertur}
\end{eqnarray}
where using the standard formuli
\begin{eqnarray}
d_n = \sum_K \left\{ C_{B1}(n,q,K) C_{B1} (n,q,K+2) + C_{B1}(n,q,K) C_{B1} (n,q,K-2)  \right\}
\label{eq:first}
\end{eqnarray}
and
\begin{eqnarray}
f_n = \sum_{j \ne n} \frac{ \left[ \sum_K \left\{ C_{B1}(n,q,K) C_{B1} (n,q,K+2) + C_{B1}(n,q,K) C_{B1} (n,q,K-2)  \right\} \right]^2  }{ (e_q^n - e_q^j) }.
\label{eq:second}
\end{eqnarray}
We limit ourselves to the regime where the lattice heights $s_1$ and $s_2$ are 
rather small $s_1, s_2 \ll 1$, and employ a simplified treatment 
to illuminate features at the crossing.
In the case of the 1st and 2nd bands, the second order perturbation term 
can be ignored, thus the lattice height
$\displaystyle{s_2^{c}=\frac{4 (e_0^2 - e_0^1) }{ ( d_2 - d_1) }}$ gives the crossing point.
As for the 3rd and 4th bands, 
the second order perturbation plays an important role, 
thus their crossing point is given by
${\displaystyle s_2^{c}=\frac{2 ( d_4 - d_3) \pm  2 \sqrt{( d_4 - d_3)^2 
-4 ( f_4 - f_3) (e_0^4 - e_0^3)} }{ ( f_4 - f_3) }}$.

\begin{figure}[htbp]
 \begin{center}
 \includegraphics[width=9cm]{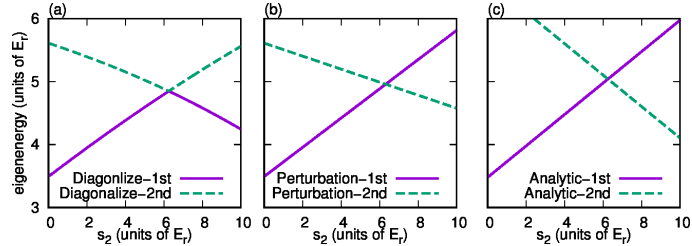}
 \end{center}
 \caption{(Color online) Eigenenergies for 1st and 2nd bands at $q=0$ as a function of $s_2$.
 In this plot the eigenenergies are shifted by $(s_1/2+s_2/2)$.
 (a), (b) and (c) show the results of exact diagonalization, perturbation theory (Eq.~\ref{eq:pertur}) and analytic function (Eq's~\ref{eq:first} and \ref{eq:second}).
 }
 \label{fig:PERT-12}
\end{figure}

First, we discuss the case of the 1st and 2nd bands.
In the limit of $s_1, s_2 \ll 1$ (see \cite{math}), the relations between the lattice height $s_1$ and the eigenenergies are given by
$e_0^1=4+\frac{s_1}{2}-\frac{s_1^2}{192}$ and $e_0^2=4+\frac{s_1}{2}+\frac{5s_1^2}{192}$.
Now suppose that dominant terms in the expansion are limited. For instance, the Bloch coefficients may be
 simply approximated by
$C_{B1}(1,0,\pm1)=\pm \frac{1}{\sqrt{2}}$ for $n=1$, $C_{B1}(2,0,\pm1)=\frac{1}{\sqrt{2}}$ for $n=2$ and 
all the other components are 0.
We thus get $d_1=-1$ and $d_2=1$ so that
\begin{eqnarray}
\tilde{E}_0^1  \simeq  4 +\frac{s_1}{2} +\frac{s_2}{2} - \frac{s_1^2}{192}  + \frac{s_2}{4} ,
\label{eq:first}
\end{eqnarray}
and
\begin{eqnarray}
\tilde{E}_0^2  \simeq  4 +\frac{s_1}{2} +\frac{s_2}{2} + \frac{5 s_1^2}{192}  - \frac{s_2}{4}.
\label{eq:second}
\end{eqnarray}
The crossing point is located at $s_2^{c}=\frac{s_1^2}{16}$.
In order to confirm the validity of this approximation at $s_1=10$, we plot the eigenenergies of the 1st and 2nd bands as a function of $s_2$ in Fig.~\ref{fig:PERT-12}.
The perturbation result agrees well with that of the exact diagonalization for small $s_2$ 
and shows the crossing.
The analytic formula of the 2nd band overestimates the eigenenergy but yields an accurate estimate
of $s_2$ for the crossing.

\begin{figure}[htbp]
 \begin{center}
 \includegraphics[width=9cm]{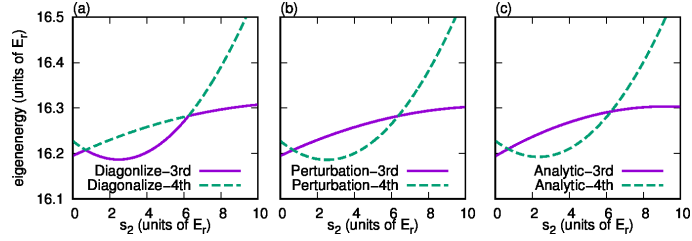}
 \end{center}
 \caption{(Color online) Eigenenergies for the 3rd and 4th bands at $q=0$ as a function of $s_2$
  as in Fig.~\ref{fig:PERT-12}.
 }
 \label{fig:PERT-34}
\end{figure}

The eigenenergies of the 3rd and 4th bands of the monochromatic OL are 
$\displaystyle{e_0^3=16+\frac{s_1}{2}+\frac{s_1^2}{480}-\frac{317s_1^4}{2^{16}3^{3}5^{3}}}$, 
$\displaystyle{e_0^4=16+\frac{s_1}{2}+\frac{s_1^2}{480}+\frac{433s_1^4}{2^{16}3^{3}5^{3}}}$ 
(see Ref.~\cite{math}).
If we apply the same simple assumption as for the $n=1$ and $2$ bands as above, we find $d_n$ goes to 0. 
Thus, we assume here that the Bloch functions have some components other then $K=\pm2$.
The Bloch coefficients for the 3rd band are then given by
$\displaystyle{C_{B1}(3,0,\pm3)=\mp \frac{40-\sqrt{40^2+s_1^2} }{\sqrt{2} s_1}}$, 
$C_{B1}(3,0,\pm2)= \pm \frac{1}{\sqrt{2}}$, 
$\displaystyle{C_{B1}(3,0,\pm1)=\pm \frac{24-\sqrt{24^2+s_1^2} }{\sqrt{2} s_1}}$ 
and the other components are 0.
For the 4th the band, the coefficients are 
$\displaystyle{C_{B1}(4,0,\pm3)= -\frac{40-\sqrt{40^2+s_1^2} }{\sqrt{2} s_1}}$, 
$\displaystyle{C_{B1}(4,0,\pm2)= \frac{1}{\sqrt{2}}}$, 
$\displaystyle{C_{B1}(4,0,\pm1)= \frac{24-\sqrt{24^2+s_1^2} }{\sqrt{2} s_1}}$,
$\displaystyle{C_{B1}(4,0,0)=\pm \frac{ (8-\sqrt{8^2+s_1^2}) (24-\sqrt{24^2+s_1^2}) }{2\sqrt{2} s_1^2}}$, 
and the other components are 0.
These values give an approximate formula 
$\displaystyle{d_4-d_3 \simeq \frac{5s_1^2}{2^8 3^2}}$.
For the second-order perturbation term, we apply the simple assumption used for the 1st and 2nd bands, such as
$\displaystyle{C_{B1}(0,0,0)=1}$, $\displaystyle{C_{B1}(3,0,\pm2)=\pm \frac{1}{\sqrt{2}}}$, 
$\displaystyle{C_{B1}(4,0,\pm2)=\frac{1}{\sqrt{2}}}$, 
$\displaystyle{C_{B1}(7,0,\pm4)=\pm \frac{1}{\sqrt{2}}}$, 
$\displaystyle{C_{B1}(8,0,\pm4)=\frac{1}{\sqrt{2}}}$, and the other components are set to 0.
The corresponding energies are $e_0^0=0$, $e_3^0=e_4^0=16$ and $e_7^0=e_8^0=64$.
This assumption gives $f_4-f_3=1/8$.
As a result, the analytic expressions are given by, 
\begin{eqnarray}
\tilde{E}_0^3  =  e_0^3 + \frac{s_2}{2} + \frac{11 s_1^2 s_2}{2^{10} 3^2 5}  - \frac{s_2^2}{2^8 3^2},
\label{eq:third}
\end{eqnarray}
and
\begin{eqnarray}
\tilde{E}_0^4  =  e_0^4 + \frac{s_2}{2} - \frac{14 s_1^2 s_2}{2^{10} 3^2 5}  + \frac{5 s_2^2}{2^8 3^2},
\label{eq:forth}
\end{eqnarray}
Finally, we obtain $s_2^{c}=\frac{5s_1^2}{2^4 3^2} \pm \frac{4s_1^2}{2^4 3^2}$.
Consequently, the first crossing point is given by $\frac{s_1^2}{144}$ and the second crossing point is given by $\frac{s_1^2}{16}$.
Fig.~\ref{fig:PERT-34} shows the $s_2$ dependence of the eigenenergies of the 3rd and 4th bands.
Again, the results of the perturbation theory and the analytic formula both show 
a good agreement with that of the exact diagonalization.
We note that the crossing points may be represented by a series of $\left( \frac{s_1}{4l} \right)^2$ where $l$ is an odd integer for $q=0$ and $\left( \frac{s_1}{4k} \right)^2$ where $k$ is an even integer for $q=\pm1$ to this order. 
It may be interesting to work out mathematical details to clarify the origin and significance of this
behavior.
We numerically checked the results up to $s_1=100$.
The mathematical features found here may be conjectured to hold up to all orders.

\end{document}